\documentclass[journal]{IEEEtran}
\ifCLASSINFOpdf
\else
   \usepackage[dvips]{graphicx}
\fi
\usepackage{url}
\usepackage{graphicx}

\ifCLASSINFOpdf
\else
\fi
\begin{document}
%
\title{Forgery Blind Inspection for Detecting Manipulations of Gel Electrophoresis Images}
%
%
%

\author{Hao-Chiang~Shao$^\dagger$,~\IEEEmembership{Member,~IEEE,}
        Ya-Jen~Cheng,
        ~Meng-Yun~Duh,
        and~Chia-Wen~Lin,~\IEEEmembership{Fellow,~IEEE}
\thanks{Prof. H.-C. Shao and Ms. M.-Y. Duh are with the Dept. Statistics and Information Science, Fu Jen Catholic University, Taiwan. Ms. M.-Y. Duh is currently chasing her master degree. (e-mail: shao.haochiang@gmail.com}
\thanks{Dr. Y.-J. Cheng is with the Office of Neuroscience Program of Academia Sinica, Institute of Molecular Biology, Academia Sinica, Taiwan. (e-mail: yajen@gate.sinica.edu.tw)}
\thanks{Prof. C.-W. Lin is with the Dept. Electrical Engineering, National Tsing Hua University, Taiwan. (e-mail: cwlin@ee.nthu.edu.tw)}%
\thanks{$^\dagger$ denotes the corresponding author.}%
\thanks{This version is an extension of Prof. Shao's previous conference paper: https://doi.org/10.1109/GlobalSIP.2018.8646594}
\thanks{Manuscript received MM-DD, YYYY; revised MM-DD, YYYY.}}

%
%

\markboth{Journal of \LaTeX\ Class Files,~Vol.~14, No.~8, August~2020}%
{Shell \MakeLowercase{\textit{et al.}}: Bare Demo of IEEEtran.cls for IEEE Journals}
%



\maketitle

\begin{abstract}
Recently, 
falsified images have been found in papers involved in research misconducts.
However, although there have been many image forgery detection methods, none of them was designed for molecular-biological experiment images. 
In this paper, we proposed a fast blind inquiry method, named FBI$_{\mbox{\tiny{GEL}}}$, for integrity of images obtained from two common sorts of molecular experiments, i.e., western blot (WB) and polymerase chain reaction (PCR). 
Based on an optimized pseudo-background capable of highlighting local residues, FBI$_{\mbox{\tiny{GEL}}}$ can reveal traceable vestiges suggesting inappropriate local modifications on WB/PCR images. Additionally, because the optimized pseudo-background is derived according to a closed-form solution, FBI$_{\mbox{\tiny{GEL}}}$ is computationally efficient and thus suitable for large scale inquiry tasks for WB/PCR image integrity.
We applied FBI$_{\mbox{\tiny{GEL}}}$ on several papers questioned by the public on \textbf{PUBPEER}, and our results show that figures of those papers indeed contain doubtful unnatural patterns.
\end{abstract}


%
\IEEEpeerreviewmaketitle

\section{Introduction}
\label{sec:intro}
\IEEEPARstart{I}{n} scientific papers, there frequently exist edited image data, resulting from inappropriately local post-processing operations---including deliberately concealed cropping of images, deliberately concealed removal of lanes from gels and blots, or excessive processing to emphasize one region in the image at the expense of others. 
Although post-processing rules have been clearly stated as guidelines or standards \cite{cellfigure,naturefigure}, some researchers still cross the borderline.
For example, as indicated by Bik et al. in 2016 \cite{bik2016prevalence}, 
about $3.8\%$ of 20,261 screened papers published in 40 journals from 1995 to 2014 contained problematic figures; 
and, many research-misconducts and faked-research scandals still happened in recent years \cite{OSUevent,kentucky2019shi,CaoEvent}.
%
To this end, 
a blind algorithm, which can detect the presence of vestige 
resulting from inappropriate local modifications on experiment images without prior knowledge of signal characteristics or imaging information, becomes a necessity.

\begin{figure}[t]
\begin{tabular}{p{200pt}}
(a) \includegraphics[width=0.43\textwidth,keepaspectratio=true]{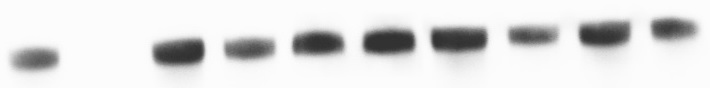}\\ 
(b) \includegraphics[width=0.43\textwidth,keepaspectratio=true]{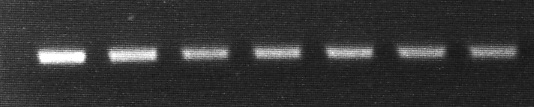} \\ 
[-0.2cm]
\end{tabular}
\caption{Example images of (a) a western blot image containing 10 bands, and (b) a gel electrophoresis PCR (polymerase chain reaction) result containing 7 band. 
}
\label{fig:01}
\end{figure}



Gel electrophoresis is a primary method for analyzing the macromolecules, e.g. DNA/RNA and proteins in the field of molecular biology. Based on the size and charge, the DNA/RNA fragments or proteins can be separated in the agarose gel. The distal images from the Western blot (WB) analysis and the gel electrophoresis result of polymerase chain reaction (PCR) sample are two common most data presented in scientific papers, as shown in Figure \ref{fig:01}. The western blot (WB) imaging, also known as protein immunoblot imaging, is an analytical technique to detect specific proteins in a sample \cite{towbin1979electrophoretic,burnette1981western}. Polymerase chain reaction (PCR) \cite{kleppe1971studies,mullis1994polymerase} is a method for making several copies of a specific DNA segment and is widely used in applications such as DNA cloning for sequencing and detection of pathogens in nucleic acid tests for the diagnosis of infectious diseases. The results of these two techniques can be recorded as digital images via a camera or a scanner and demonstrated as grayscale images. 
Because the whole imaging process is complex and time-consuming, people may dishonestly create new results without repeating the experiments by editing/modifying existed WB/PCR images. 
The most common post-processing operation of WB/PCR imagery is cloning, i.e., copy-paste. Hence, although there have been several categories of image forgery detection methods \cite{farid2009image,birajdar2013digital,qazi2013survey,pun2015image}, only copy-move forgery detectors and feature-based methods, such as \cite{christlein2012evaluation,pun2015image,zandi2016iterative} and \cite{li2017image}, are related to this issue. 
However, because a user can always copy a WB/PCR band from one unpublished experiment result and paste it onto another, it is challenging to discriminate whether an unseen WB/PCR band is a clone or not. 
Moreover, even if two bands are suspected to be clones from the same data
due to a high PSNR (peak signal-to-noise ratio) value between them, 
it is still not sufficient to rule out the possibility that these two bands are just similar to each other. 
Consequently, how to reveal the vestige of inappropriate post-processing operations, if any, becomes the prior concern in this molecular-biological image forgery detection problem.





The proposed FBI$_{\mbox{\tiny{GEL}}}$ aims to highlight the discontinuity, caused probably by man-made modifications, in the magnitude of estimated background noise. 
Because each scanning/photoing procedure ought to have its own noise pattern and noise distribution, 
post-processing operations would hopefully leave some traceable vestiges, i.e., \textit{unnatural patterns}, on a modified image. 
Specifically, because natural images usually have curvilinear contours and smooth transition areas, we can detect and even localize modifications by checking if there exist i) discontinuity or clear-cut boundaries in the background noise, and ii) rectangular contours on an input WB/PCR image. 
Consequently, FBI$_{\mbox{\tiny{GEL}}}$ follows a common idea adopted in the TFT-LCD mura defect detection problem for estimating a pseudo-background of a very low-contrast image and adopts a classical concept used in Macro-Economics, i.e., Hodrick-Prescott filter \cite{hpfilter1997}, for separating the volatility term from a 1D time series. 
We design an optimization equation and derive its closed-form solution to estimate a best suitable background trend and the residue component of an input image. Based on this design, not only we can quantitatively describe the condition for revealing unnatural patterns, but also the optimized result can be derived in a deterministic fashion rather than a stochastic manner.
%
%
In sum, FBI$_{\mbox{\tiny{GEL}}}$ performs blind detection by making invisible unnatural patterns just noticeable, as the concept described in \cite{SSO,watson2007spatial,shao2009robust}. 






The contributions of FBI$_{\mbox{\tiny{GEL}}}$ are therefore threefold.\\  
    \noindent $\bullet$ 
    Extending background analysis to a new realm, 
    FBI$_{\mbox{\tiny{GEL}}}$ is the first blind inquiry method for  integrity of gel electrophoresis images. \\
    \noindent $\bullet$ FBI$_{\mbox{\tiny{GEL}}}$ is nearly parameter free, and all its inspection results are derived on the same basis. Therefore, FBM$_{\mbox{\tiny{GEL}}}$ can avoid false-alarms effectively, as will be described in Section \ref{sec:exp}. \\
    \noindent $\bullet$ FBI$_{\mbox{\tiny{GEL}}}$ operates based on the closed-form solution described in Eq. (\ref{eq06:sol}). It is computationally fast and thus suitable for large-scale inquiry tasks. 
\section{Related Work}
\label{sec:review}
Research integrity is the most essential thing in our research community. 
In order to deter tampering with experiment images, the Office of Research Integrity of U.S. already provided a photoshop-based macro, named \textit{Droplets} \cite{droplets}, for screening falsified WB/PCR images. However, because \textit{Droplets} operates based on only histogram equalization, gradient map and pseudo-coloring, it will fail to reveal features if i) the inspector sets inappropriate parameters, or ii) the brightness of a falsified WB/PCR image was carefully adjusted. Hence, \textit{Droplets} cannot be effective enough. 

Revealing invisible unnatural patterns on WB/PCR image is conceptually similar to detecting 
``mura defect'' that describes the uneven patches of changes in luminance on thin-film-transistor liquid-crystal display (TFT-LCD) panels. 
However, most mura detectors were designed to identity defect regions, in which brightness is slightly different from the low-contrast background. Mura detection methods, such as \cite{watson2007spatial,shao2009robust,watson2005standard,watson2000visual,lee2004automatic,taniguchi2006mura}, would be severely disturbed by foreground bands of WB/PCR images and thus not suitable for WB/PCR forgery detection.

Therefore, we developed our FBI$_{\mbox{\tiny{GEL}}}$ in a very conservative way by i) following the common idea of pseudo-background estimation used in mura defector design, and ii) exploiting a classical concept, named Hodrick-Prescott (HP) filter \cite{hpfilter1997}, in Macroeconomics. 
The HP filter was developed to separate 
permanent shocks $\tau_t$, i.e., the stationary component or the trend component, of a 1D data series, which depicts a business cycle, from the temporary shocks $c_t = y_t - \tau_t$ causing volatility, i.e., the non-stationary part or cyclic component. 
The concept of HP filter can be summarized as below
\begin{equation}
\min_{\tau_t} \Sigma^T_{t=1}(y_t-\tau_t)^2 + \lambda \Sigma^{T-1}_{t=2} 
[ (\tau_{t+1} - \tau_t) - (\tau_t - \tau_{t-1}) ]^2 \mbox{.} 
\label{eq01:HP}
\end{equation}

\section{Fast Blind Inquiry for Integrity of Gel Electrophoresis Imagery}
\label{sec:method}


%
%
%


Similar to mura detection methods designed for low-contrast images, the proposed method needs to extract an estimated pseudo-background from an input WB/PCR image first. 
Inspired by a general idea of extracting trend and volatility components within a 1-dimensional time series, we use the optimization equation shown in Eq. (\ref{eq02:HP2d}) to define a pseudo-background $\mathcal{J}$ of an input image $\mathcal{I}$.
That is, 
\begin{equation}
\min_\mathcal{J} \parallel \mathcal{I}-\mathcal{J} \parallel^2_F + \lambda \parallel h * \mathcal{J} \parallel ^2_F 
\mbox{,}
\label{eq02:HP2d}
\end{equation}
where $h$ denotes a high pass kernel.
This equation forces the to-be-estimated trend component $\mathcal{J}$ to be a smoothened approximation of $\mathcal{I}$. Obviously, when $\mathcal{J}$ and $\mathcal{I}$ be images of dimension $M \times 1$ and $h=[1,-2,1]$, Eq. (\ref{eq02:HP2d}) degenerates to the 1D Hodrick-Prescott filter described in Eq. (\ref{eq01:HP}). 

However, one primary concern, in biologists' points of view, in this blind inspection problem is whether the solution can be deterministic. 
Any randomness in the optimization solver may make it difficult for the investigation committee to role-out possible chance coincidences.
Hence, a closed-form solution of Eq.(\ref{eq02:HP2d}) is required. To find the closed-form solution, we rewrote first Eq. (\ref{eq02:HP2d}) 
by replacing Frobenius norm by matrix trace. That is, 
\begin{eqnarray}
\mathcal{L} &=& 
\parallel \mathcal{I}-\mathcal{J} \parallel^2_F + \lambda \parallel h * \mathcal{J} \parallel ^2_F \nonumber \\
&=& tr \{ (\mathcal{I}-\mathcal{J}) (\mathcal{I}-\mathcal{J})^t \} 
+ tr\{ H \mathcal{J} H^t H \mathcal{J}^t H \}
\mbox{,}
\label{eq03:derivesol}
\end{eqnarray}
where $H$ is the Toeplitz matrix of a 1D $k$-tap-long high-pass filter $f_{k \times 1}$. That is, $h$ is now assumed to be separable and $h=f\, f^t$, where ``$t$'' denotes matrix transpose.
Then, letting the partial derivative of $\mathcal{L}$ with respect to $\mathcal{J}$ be zero, we obtain
\begin{eqnarray}
\frac{\partial}{\partial \mathcal{J}}\mathcal{L} 
&=& 2 \mathcal{J}^t - 2 \mathcal{I}^t + 2 \lambda (H^t H \mathcal{J}^t H^t H)=0 \mbox{.}
\label{eq04:derivesol}
\end{eqnarray}
Therefore, we have 
\begin{equation}
\mathcal{I} = \mathcal{J} + \lambda (H^t H \mathcal{J} H^t H) \mbox{.}
\label{eq05a:solsrc}
\end{equation}
This equation guarantees that the input image $\mathcal{I}$ can be expressed as a weighting sum of a low-passed component $\mathcal{J}$ and a high-passed term $H^t H \mathcal{J} H^t H$.

Because $H \mathcal{J} H^t$ denotes the matrix multiplication form of a 2D  convolution $h*\mathcal{J}$, Eq. (\ref{eq05a:solsrc}) can be rewritten as 
\begin{equation}
\mathcal{I} = \mathcal{J} + \lambda ( h^t * (h* \mathcal{J}) )
\mbox{.}
\label{eq05c:solsrc}
\end{equation}
Consequently, by using discrete Fourier transform, the closed-form solution of $\mathcal{J}$ can be derived as 
\begin{equation}
\mathcal{J} = \mathcal{F}^{-1} \left \{
\frac{\mathcal{F}\{\mathcal{I}\}}
{  \mathbf{[1]} + \lambda \mathcal{F}\{h^t \} \circ \mathcal{F}\{h\} }  
\right \} \mbox{,}
\label{eq06:sol}
\end{equation}
where $\circ$ denotes Hadamard product, and $\mathbf{[1]}$ is a constant matrix whose entries are all 1's.
Based on $\mathcal{J}$, the residue pattern $\mathcal{E}$ of the input image $\mathcal{I}$ is defined to be the absolute difference between $\mathcal{I}$ and $\mathcal{J}$, i.e., $\mathcal{E}=| \mathcal{I}-\mathcal{J} |$. Based on $\mathcal{E}$, the vestige of man-made modifications can be revealed.




\subsection{Selecting A High-pass Filter $h$}
\label{subsec:301}

Although $h$ was assumed to be separable whiling deriving the closed-form solution, this constraint can be relaxed in practice. Based on the concept described by Eq.(\ref{eq05c:solsrc}) that the input image $\mathcal{I}$ is a linear combination of a low-passed component $\mathcal{J}$ and a high-passed term $h^t * (h* \mathcal{J})$ extracted from $\mathcal{J}$, any high-pass filter $h$, no matter it is separable or not, can result in an optimal pseudo-background $\mathcal{J}(h, \lambda)$. 
In practice, we set $h = \delta - f$, where $f$ is a Gaussian kernel, e.g. a kernel given by MATLAB built-in function \textbf{fspecial(`gaussian', [3 3], 1.0)}. 


\subsection{Revealing Unnatural Patterns}
\label{subsec:302}

To check if there exists any unusual pattern, the obtained $\mathcal{E}$ is further processed by following steps: \\
\textbf{Step-1:} Bring all pixel values of $\mathcal{E}$ into the range $[0, 1]$ and apply hard thresholding on normalized $\mathcal{E}$ with a user-specified threshold value $\gamma$. \\
\textbf{Step-2:} Binarize the thresholded normalized $\mathcal{E}$ to obtain an indicator map $\mathcal{M}_{\lambda, \gamma}$, as examples shown in Figures \ref{fig:exp04} and \ref{fig:exp01}. \\
%
%
%
%
\textbf{Step-3:} Fuse $\mathcal{M}_{\lambda, \gamma}$ and $\mathcal{I}$ via alpha blending, after staining the white area of $\mathcal{M}_{\lambda, \gamma}$ yellow, to highlight where unnatural patterns locate, as the example shown in Figure \ref{fig:exp01}(c). \\
Empirically, the default value of $\lambda$ is $0.00005$; and, the default value of $\gamma$ is $0.0001$, although any value in between $0.0001$ and $0.5$ is good for $\gamma$.

%

\section{Experiment Result}
\label{sec:exp}

\subsection{Robustness against compression}
\label{subsec:403b}
\begin{figure}[!t]
\begin{tabular}{p{200pt}}
\includegraphics[width=0.48\textwidth, height=10cm,keepaspectratio=true]{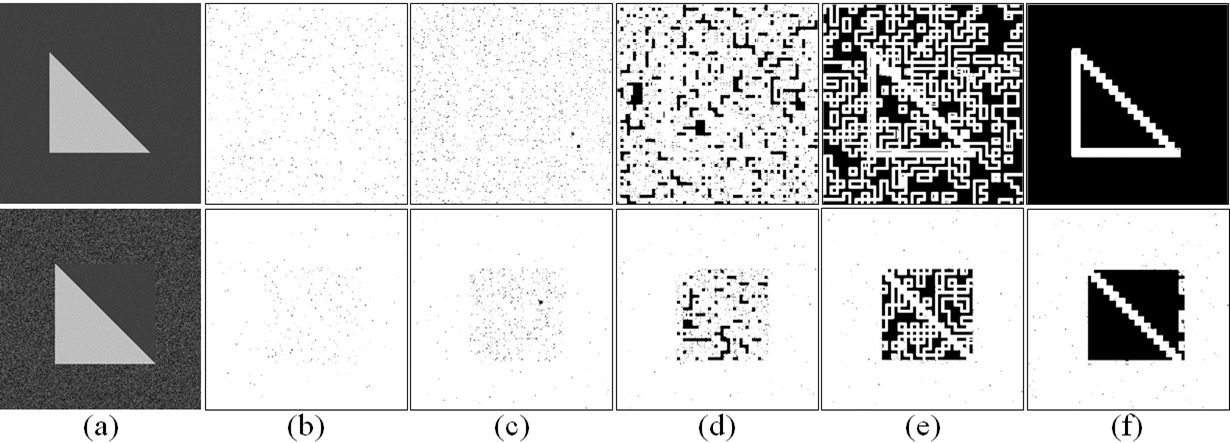}
\end{tabular}
\caption{Comparisons among inquiry results of sample patterns compressed with different quality settings. Leftmost column: source images. Columns (b)--(f): indicator maps of sample patterns compressed under quality value. 
Top row: simulations of unmodified WB/PCR images. Bottom row: simulation of copy-pasted WB/PCR images.
Note that we used PhotoShop to produce JPEG versions of source mother images in BMP format. The compression quality of column-(b), -(c), -(d), -(e), and -(f) were set to be 10, 7, 5, 3, and 1, respectively.
}
\label{fig:exp04}
\end{figure}

We first verify FBI$_{\mbox{\tiny{GEL}}}$'s robustness against JPEG compression because published images are all compressed. 
In Figure \ref{fig:exp04}, column-(a) shows mother images, and each of other columns is associated with a different compression quality setting. 
The upper row of Figure \ref{fig:exp04} shows inquiry results of simulated unmodified WB/PCR images whose foreground, i.e., the triangle and its bounding box, and background were contaminated by the same level of Gaussian noise ($\sigma=0.01$); and, the lower row contains results of copy-pasted forgery simulations whose background and foreground components were affected by different levels of Gaussian noises ($\sigma_{\mbox{\small{back}}}=0.1$ and $\sigma_{\mbox{\small{fore}}}=0.01$). 
%
%

Based on Figure \ref{fig:exp04}, we have four concluding remarks. First, for an unmodified image with homogeneous noise (upper row), the indicator map $\mathcal{M}
_{\lambda, \gamma}$ shows the same patterns on both background and foreground, no matter how severely an image was compressed. Second, for an edited WB/PCR images (lower row), the indicator map can highlight the difference between background template and copy-pasted foreground. Third, a high compression ratio (low compression quality) will not invalidate FBI$_{\mbox{ \tiny{GEL}}}$; instead, a high compression ratio can make the copy-pasted foreground more distinguishable from background template. Fourth, the most important of all, by conservatively reporting only the pattern shown in the lower part of Figure \ref{fig:exp04}(f) as an almost-surely falsification, FBI$_{\mbox{\tiny{GEL}}}$ is expected to be capable of avoiding false-alarm. 

\begin{figure}[!t]
\begin{tabular}{p{200pt}}
(a) \includegraphics[width=0.44\textwidth, height=10cm,keepaspectratio=true]{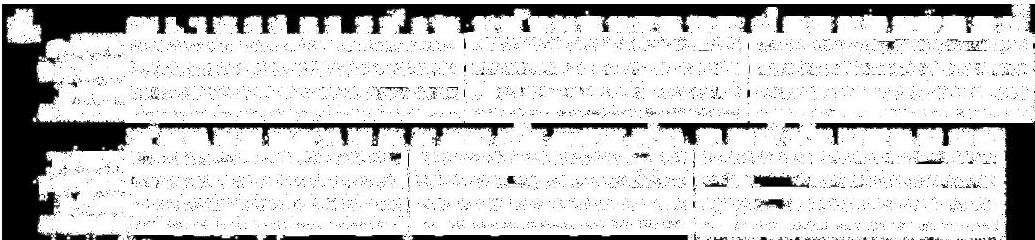} \\
(b) \includegraphics[width=0.44\textwidth, height=10cm,keepaspectratio=true]{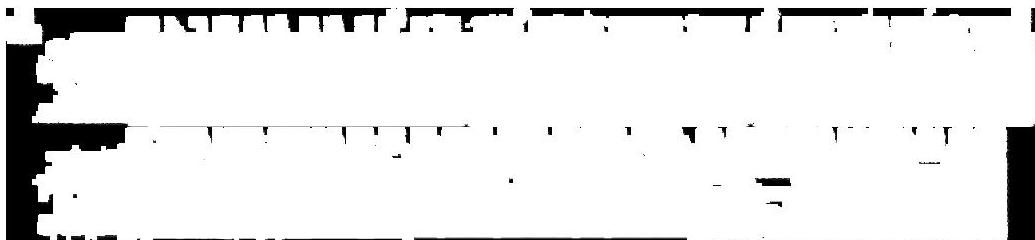}\\
(c) \includegraphics[width=0.44\textwidth, height=10cm,keepaspectratio=true]{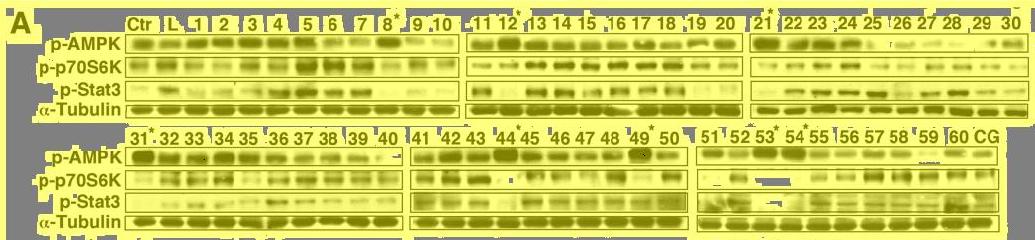} \\
 [-0.2cm]
\end{tabular}
\caption{ Inquiry results of \textbf{Figure-4A in \cite{guh2010development}}. 
(a) $\mathcal{M}_{0.00005, 0.5}$; (b) 
$\mathcal{M}_{0.00005, 0.0001}$; and, (c) the alpha blending result of the input image $\mathcal{I}$ and the stained $\mathcal{M}_{0.00005, 0.0001}$.}
\label{fig:exp01}
\end{figure}

\subsection{Tests on Open Data}
\label{subsec:exp01}




Figure \ref{fig:exp01} shows the analysis results of \textbf{Figure-4A in \cite{guh2010development}}, one of the retracted papers listed in the investigation report released by Ohio State University \cite{OSUevent,Osureport2}. 
There are 252 western blot (WB) image bands---4 bands of \textbf{Ctr}, 4 bands of \textbf{L}, 4 bands of \textbf{CG}, and other $4*60$ bands---in Figure \ref{fig:exp01}(c); however, both the indicator map $\mathcal{M}_{\lambda, \gamma}$ demonstrated in Figure \ref{fig:exp01}(a)-(b) show that the $53^{rd}$ band in the row entitled \textbf{p-p70S6K} is an empty zone (the horizontal black stripe). In addition, we can also observe that 
although different $\gamma$ values result in different indicator maps, locations and contour shapes of empty zones on Figures \ref{fig:exp01}(a)-(b) are nearly consistent. Consequently, we deem that this empty zone forms an unnatural pattern, which might be a vestige of man-made modifications, such as erasure or block-wise region removal. 

\begin{figure}[!b]
\begin{tabular}{p{200pt}}
(a) \includegraphics[width=0.45\textwidth, keepaspectratio=true]{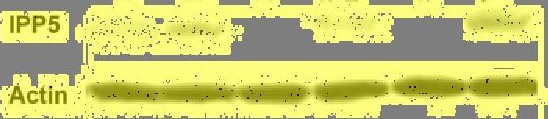}\\ 
(b) \includegraphics[width=0.45\textwidth, keepaspectratio=true]{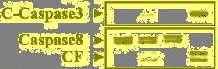}\\
(c) \includegraphics[width=0.45\textwidth, keepaspectratio=true]{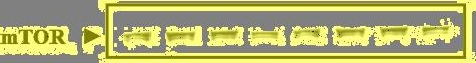} \\
(d) \includegraphics[width=0.45\textwidth, keepaspectratio=true]{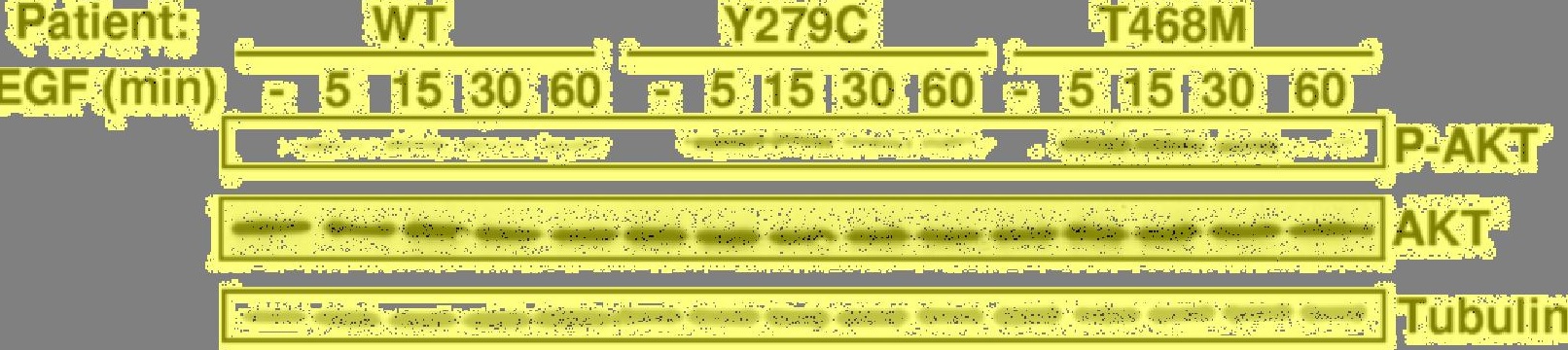} \\ [-0.2cm]
\end{tabular}
\caption{Inquiry results of figures questioned by the public. 
(a) result of \textbf{Figure-1C in \cite{wang2008ipp5}}; 
(b) result of \textbf{Figure-6D in \cite{gao2009induction}};
(c) result of \textbf{Figure-3 in \cite{gao2009induction}}; and, 
(d) result of \textbf{Figure-1B in \cite{edouard2010functional}}.
}
\label{fig:exp02}
\end{figure}

Next, demonstrated in Figure \ref{fig:exp02} are the WB images published with \cite{edouard2010functional,wang2008ipp5} and \cite{gao2009induction}; these three papers have already drawn public attention and been questioned on following \textbf{Pubpeer} discussion pages: \cite{edouard2010PUBPEER,wang2008PUBPEER,gao2009PUBPEER}.
Here, three types of patterns can be observed. 
Examples of type-1 pattern include the \textbf{$\mathbf{\alpha}$-Tubulin} and \textbf{p-AMPK} rows in Figure \ref{fig:exp01}(c), and the AKT and the Tubulin rows in Figure \ref{fig:exp02}(d).
We consider this kind of patterns is normal and standard because its indicator map  $\mathcal{M}_{\lambda, \gamma}$ is homogeneous and contains no empty zone and no vertical interruption stripe. 
Furthermore, we define the type-2 pattern to be an almost empty zone. For example, the $3^{rd}$ and the $5^{th}$ bands of the \textbf{IPP5} row in Figure \ref{fig:exp02}(a), the $3^{rd}$ band of the \textbf{C-Caspase3} row and the $4^{th}$ band of the \textbf{Caspase8} row in Figure \ref{fig:exp02}(b), and the three bands (of the \textbf{P-AKT} row) beneath the minus signs in Figure \ref{fig:exp02}(d) are all of this kind. 
We consider the type-2 pattern unnatural and surmise that this pattern might result from a procedure similar to the one causes the empty zone demonstrated in Figure \ref{fig:exp01}. 
At the end of this paragraph, we need to emphasize that the $2^{nd}$ band of the \textbf{C-Caspase3} row in Figure \ref{fig:exp02}(b) has a sharp vertical edge at its right-hand-side boarder, and this appearance must not be a change coincidence.  

Finally, the type-3 pattern includes those containing block-wise non-empty zones and those containing stripe-wise empty zones. For instance, bands in the \textbf{mTOR} row in Figure \ref{fig:exp02}(c) and the $2^{nd}$ band in the \textbf{CF} row in Figure \ref{fig:exp02}(b) are exactly of this kind. 
Primary features of a type-3 pattern include i) a band (or an area containing multiple bands) that is independently surrounded with a rectangular zone formed by non-zero entries of the indicator map $\mathcal{M}_{\lambda, \gamma}$, and ii) a non-zero zone that is secluded from each other by some narrow, vertical stripe-wise empty zones (gray area). 
Because it does not make sense to modify the gel background where no reaction/response happens, we conjectured that a type-3 pattern would result from a copy-paste of a rectangular region. 
Such conjecture is supported by the fact that, by using template matching, the PSNR between the $2^{nd}$ and the $6^{th}$ bands of the \textbf{mTOR} row in Figure \ref{fig:exp02}(c) is 23.81 dB, and the PSNR between the $3^{rd}$ and the $7^{th}$ bands is 22.08 dB. 
Besides, because the public also suspected that the \textbf{mTOR} row contains copy-move forgeries \cite{wang2008PUBPEER}, we exploited following experiments to clarify this situation (For more demonstrations, please refer to our supplemental document and \cite{Shao2018Unveiling}.). 


%

%
%

\subsection{Interpretation of type-2 and type-3 patterns:}
\label{subsec:exp02}


We utilized one additional simulation dataset to clarify the cause of each unnatural pattern we met. This simulation dataset consists of thirteen post-processed WB/PCR images, which were designed to reproduce the unnatural patterns demonstrated in previous subsection. 

\begin{figure}[t]
\begin{tabular}{p{200pt}}
(a) \includegraphics[width=0.44\textwidth, height=10cm,keepaspectratio=true]{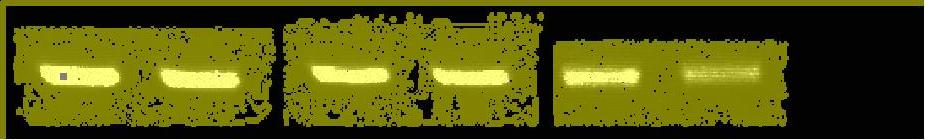}
(b) \includegraphics[width=0.44\textwidth, height=10cm,keepaspectratio=true]{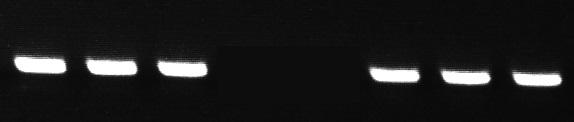}
(c) \includegraphics[width=0.44\textwidth, height=10cm,keepaspectratio=true]{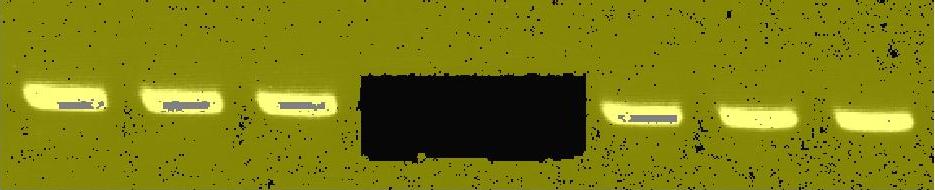} 
\end{tabular}
\caption{Experiments on the control group. (b): images with man-made modifications. (a) and (c): inquiry results.}
\label{fig:exp03}
\end{figure}

%
%
%
%
%
Figures \ref{fig:01}(b) and \ref{fig:exp03}(b) were designed to clarify the causes of type-2 and type-3 patterns. Figure \ref{fig:01}(b) was created by copying three rectangular areas independently from other PCR images, pasting them together on the same template, and then adjusting image brightness and contrast properly. Meanwhile, Figure \ref{fig:exp03}(b) was created by removing its $4^{th}$ and $5^{th}$ bands from the source image. 
Hence, Figure \ref{fig:exp03}(b) is disguised as a new PCR result with negative response at its $4^{th}$ and $5^{th}$ bands, and Figure \ref{fig:01}(b) looks as if a common experiment result containing six positive bands and one negative band.
%
%

Demonstrated in Figures \ref{fig:exp03}(a) and (c) are the inquiry results derived by FBI$_{\mbox{\tiny{GEL}}}$. The black rectangular region in Figure \ref{fig:exp03}(c) denotes an empty zone in the indicator map, and this empty zone corresponds to what we removed from the source image. Figure \ref{fig:exp03}(a) also reveals the way we create Figure \ref{fig:01}(b)---a background template with three copy-pasted rectangular foreground. 
Consequently, a possible way to create type-2 pattern is erasure or block-wise region removal, and the type-3 pattern can be reproduced by a typical copy-paste procedure. FBI$_{\mbox{\tiny{GEL}}}$ also confirmed that the questions raised by the public on \textbf{PUBPEER} were reasonable, and images on those papers were indeed problematic.

\section{Concluding Remarks}
\label{sec:conclu}
In this paper, we proposed a fast blind inquiry method, named FBI$_{\mbox{\tiny{GEL}}}$, for integrity of images obtained 
from western blot (WB) and polymerase chain reaction (PCR) results. 
Based on an optimized pseudo-background, FBI$_{\mbox{\tiny{GEL}}}$ can reveal traceable vestiges of inappropriate local modifications on WB/PCR images. Also, FBI$_{\mbox{\tiny{GEL}}}$ is suitable for large scale inspection tasks for WB/PCR image integrity because it is computationally efficient. Our experiment results show that images on papers questioned by the public on \textbf{PUBPEER} are indeed doubtful.  
Finally, we have to emphasize two points. First, FBI$_{\mbox{\tiny{GEL}}}$ was not designed for accusing anyone; instead, FBI$_{\mbox{\tiny{GEL}}}$ was developed for helping academic community identify problematic figures and irreproducible experiments. Second, whether an image with unnatural pattern invalidates 
its significance in that very field is beyond the scope of this FBI$_{\mbox{\tiny{GEL}}}$ method.

\section*{Acknowledgement}
This work is supported by the Ministry of Science and Technology, Taiwan (MOST 107-2320-B-030-012-MY3). The author wants to thank Prof. Cheng-Ting Chien for providing source WB/PCR images. 
The authors also want to thank Prof. Yung-Chang Chen and Dr. Hsiu-Ming Chang for their suggestions on this paper.

\ifCLASSOPTIONcaptionsoff
  \newpage
\fi



%

\bibliographystyle{IEEEtran}
\bibliography{refs_biomolefake}

%
%

%

\begin{IEEEbiography}
[{\includegraphics[width=1in,height=1.25in,clip,keepaspectratio]{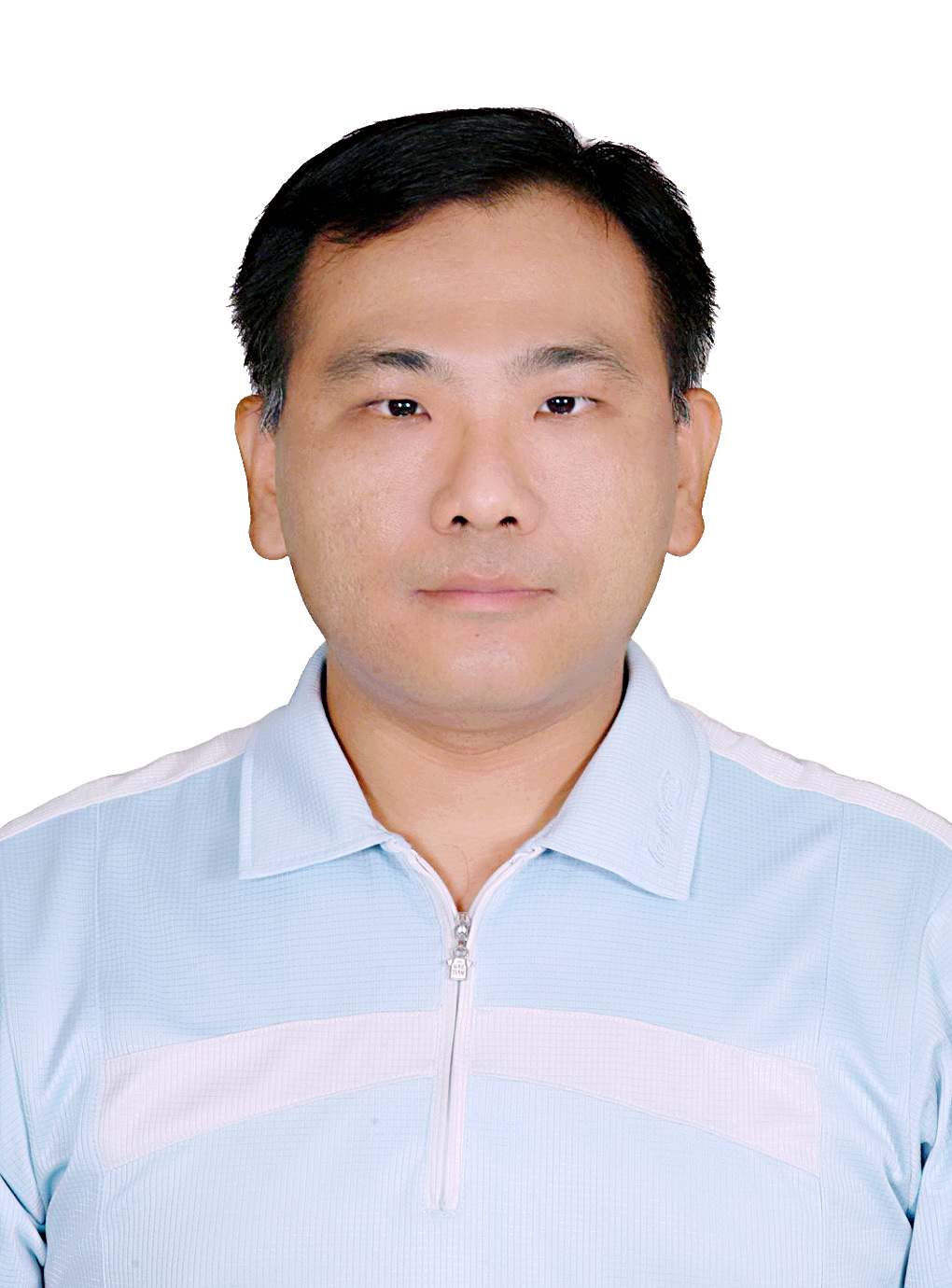}}]
{Hao-Chiang~Shao}
(Member, IEEE) received his Ph.D. in electrical engineering from National Tsing Hua University, Taiwan, in 2012. He has been an Assistant Professor with the Dept. Statistics and Information Science, Fu Jen Catholic University, Taiwan, since 2018. During 2012 to 2017, he was a postdoctoral researcher with the Institute of Information Science, Academia Sinica, involved in a series of \textit{Drosophila} brain research projects; in 2017--2018, he was an R\&D engineer with the Computational Intelligence Technology Center, Industrial Technology Research Institute, Taiwan, taking charges of DNN-based automated optical inspection (AOI) projects. His research interests include 2D+Z image atlasing, 3D mesh processing, big industrial image data analysis, and machine learning.
\end{IEEEbiography}

\begin{IEEEbiographynophoto}{Ya-Jen~Cheng}
received her Ph.D. in molecular and cellular biology from National Tsing Hua University, Taiwan, in 2012. She has been the manager to coordinate and execute the schemes planed by Neuroscience Program of Academia Sinica (NPAS) since 2010. She experts in creating strategic management, setting budgets, and managing risks with the leader and team members to effectively align with and support key research and academic expectations. With her research background in genetics and imaging techniques, she is also the manager of the imaging equipment and the related analyzing system in the neuroscience core facility. She is responsible for arranging the technical training and also training for imaging processing.
\end{IEEEbiographynophoto}


\begin{IEEEbiographynophoto}{Meng-Yun~Duh}
received her B.A. from Fu Jen Catholic University, Taiwan, in 2019. Her research interests include statistics, data science, and biomedical image forgery detection. She is now chasing her M.A. degree.
\end{IEEEbiographynophoto}

\begin{IEEEbiography}
[{\includegraphics[width=1in,height=1.25in,clip,keepaspectratio]{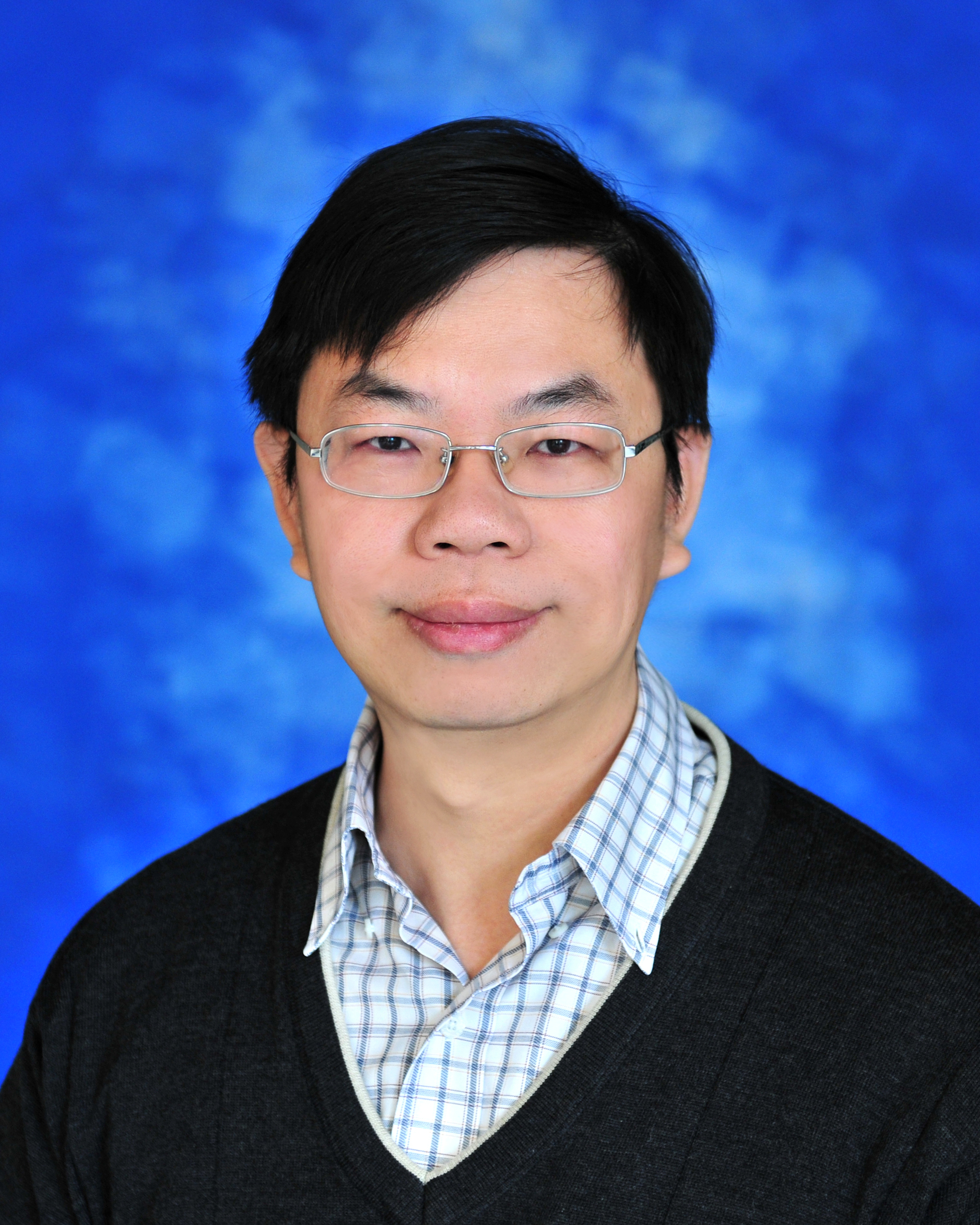}}]{Chia-Wen~Lin}
(Fellow, IEEE) received his Ph.D. from National Tsing Hua University (NTHU), Taiwan, in 2000.  
	Dr. Lin is currently Professor with the Department of Electrical Engineering and the Institute of Communications Engineering, NTHU.   His research interests include image/video processing, computer vision, and machine learning.  He served as Distinguished Lecturer of IEEE Circuits and Systems Society (2018--2019).   He is Chair of IEEE ICME Steering Committee. He served as TPC Co-Chair of IEEE ICIP 2019 and IEEE ICME 2010, and General Co-Chair of IEEE VCIP 2018. He was a recipient of Outstanding Electrical Engineer Professor Award presented by the Chinese Institute of Electrical Engineering, Taiwan. He received two best paper awards from VCIP 2010 and 2015. He has served as an Associate Editor of \textsc{IEEE Transactions on Image Processing}, \textsc{IEEE Transactions on Circuits and Systems for Video Technology}, \textsc{IEEE Transactions on Multimedia}, and \textsc{IEEE Multimedia}.  He served as a Steering Committee member of \textsc{IEEE Transactions on Multimedia} from 2013 to 2015.
\end{IEEEbiography}

\end{document}